\documentclass[3p,twocolumn,sort&compress, fleqn]{elsarticle}
\usepackage{amsmath, bm}
\usepackage{amssymb}
\usepackage{braket}
\journal{Computer Physics Communications}

\def\G1{{G^{(1)}}}

\usepackage{natbib}
\usepackage{hyperref}
\hypersetup{
        colorlinks=true,
}
\usepackage{listings}

\long\def\beginmypgfpdfnamed#1#2\endmypgfpdfnamed{\includegraphics{#1}}

\graphicspath{{./pyfig/}}

\begin{document}

\begin{frontmatter}

\title{
	Efficient implementation of the continuous-time interaction-expansion quantum Monte Carlo method
}

\author[saitama]{Hiroshi Shinaoka} \ead{shinaoka@mail.saitama-u.ac.jp}
\author[univtokyo]{Yusuke Nomura}
\author[michigan]{Emanuel Gull}

\address[saitama]{Department of Physics, Saitama University, Saitama 338-8570, Japan}
\address[univtokyo]{Department of Applied Physics, University of Tokyo, 7-3-1 Hongo, Bunkyo-ku, Tokyo, 113-8656, Japan}
\address[michigan]{Department of Physics, University of Michigan, Ann Arbor, MI 48109, USA}

\begin{abstract}
We describe an open-source implementation of the continuous-time interaction-expansion quantum Monte Carlo method for cluster-type impurity models with onsite Coulomb interactions and complex Weiss functions. 
The code is based on the ALPS libraries.
\end{abstract}


\end{frontmatter}
{\bf PROGRAM SUMMARY}

\begin{small}
\noindent
{\em Program Title:} ALPS CT-INT \\
{\em Journal Reference:}                                      \\
{\em Catalogue identifier:}                                   \\
{\em Licensing provisions:} GPLv3
{\em Programming language:} \verb*#C++#, MPI for parallelization. \\
{\em Computer:} PC, HPC cluster \\ 
{\em Operating system:} Any, tested on Linux and Mac OS X\\ 
{\em RAM:} 100 MB - 1 GB.\\ 
{\em Number of processors used:} 1 - 2000.
{\em Classification:} 4.4  \\ 
{\em External routines/libraries:}  ALPSCore libraries, Eigen3, Boost.\\ 
{\em Nature of problem:} Quantum impurity problem\\
{\em Solution method:} Continuous-time interaction expansion quantum Monte Carlo\\
{\em Running time:} 1 min -- 8 h (strongly depends on the problem to solve)\\
\end{small}

\section{Introduction}
Quantum impurity problems describe small  interacting  sets of orbitals coupled to wide non-interacting leads. 
Originally developed in the context of magnetic impurity atoms embedded in a non-magnetic host \cite{Anderson61}, they have since found applications to quantum dots and molecular conductors \cite{Hanson07} , atoms adsorbed to surfaces \cite{Brako81}, and appear as auxiliary objects in quantum embedding theories such as the dynamical mean field theory \cite{Georges:1996un},  its extensions \cite{Hettler:1999jj,Lichtenstein:2000bp,Maier:2005et,Anonymous:lduB2q0_,Rubtsov:2008cs} and the self-energy embedding theory \cite{Kananenka15,Zgid17}.

Most of these applications require the calculation of impurity model energies, green's functions, and self-energies in a non-perturbative regime.  Analytic methods are ill suited to this task, and one needs to resort to numerical methods such as the numerical renormalization group \cite{Bulla:2008bn}, exact diagonalization \cite{Caffarel:1994cv},  configuration interaction \cite{Zgid:2012ck}, density matrix renormalization group theory \cite{Garcia04,Nishimoto06,Wolf14}, or quantum Monte Carlo \cite{Gull:2011jda,Rubtsov:2003vb,Rubtsov:2005iwa,Werner:2006ko,Werner:2006iz,Gull:2008cma}.

Many embedding methods, especially when formulated as cluster theories \cite{Maier:2005et} for simplified low-energy effective models, generate impurity models that have general off-diagonal and potentially complex-valued hybridization functions but interactions of the density-density type. Models with off-diagonal complex hybridization functions are substantially more difficult to solve than those with diagonal hybridizations, and require the use of specialized impurity solvers.

In this paper, we describe an open source implementation of such a solver. The method is an implementation of the algorithm developed by Rubtsov {\it et al.} \cite{Rubtsov:2005iwa} and implements the stochastic sampling of a weak coupling perturbation series to all orders. The algorithm also implements the submatrix update scheme of Refs.~\cite{Nukala:2009gh,Gull:2011hh,Nomura:2014by}. In the absence of a sign problem, it scales cubically as a function of system size, inverse temperature, and interaction strength. In general, results at low temperature are hampered by an exponential scaling do to a fermionic sign problem \cite{Shinaoka:2015wq}.

The remainder of this paper is organized as follows. In section \ref{sec:model_and_algorithm} we introduce the model and algorithm. In section \ref{sec:usage} we show the usage of the code. In section \ref{sec:examples} we illustrate the usage of the code at a few examples. Section \ref{sec:summary} contains our conclusions.

\section{Model and algorithm}\label{sec:model_and_algorithm}
The current version of the ALPS/CT-INT impurity solver supports 
single-orbital multi-site ($N$-site) impurity models with onsite Hubbard interactions 
defined by the action
\begin{eqnarray}
	S_{\rm{imp}} =  S_0 + S_{\rm{int}},
\end{eqnarray}
where
\begin{align}
	S_0 = - \iint^{\beta}_{0}  d{\tau} d{\tau'} \sum_{i,j=0}^{N_s -1 }  \sum_{\sigma=0}^{N_\sigma-1}
 &	\bigl [ \mathcal{G} ^{-1}_{0\sigma} (\tau - \tau^\prime) \bigr]_{ij}  \times \nonumber \\ 
 &	 \hat{c}^{\dagger}_{i\sigma}(\tau) \hat{c}^{\ }_{j\sigma} (\tau^\prime), \nonumber 
\end{align}
where $\sigma$, $i$, and $j$ are indices and the double integration goes over $\tau$ and $\tau'$.
Here, $\hat{c}_{{i\sigma}}^{\dagger}$ ($\hat{c}_{i\sigma}$) is a Grassmann variable representing the creation (annihilation) of an impurity electron specified by indicies $i$ and $\sigma$.
The solver assumes the Weiss function $\mathcal{G} ^{-1}_{0\sigma} (\tau - \tau')$ to be diagonal in $\sigma$ but it can be off-diagonal in $i$ and $j$.
The Weiss function is a (complex-valued) matrix with respect to the index $i$ and $j$ 
for each $\sigma$.

In a typical single-orbital multi-site impurity model with two spins $|\uparrow\rangle$ and $|\downarrow\rangle$ but no hopping between different spins, $\sigma$ enumerates spins ($N_\sigma=2$) and $i$ and $j$ the $N$ impurity sites ($N_s=N$).
The interaction part 
is defined as 
\begin{align}
S_{\rm{int}} &= \int^{\beta}_{0}  d \tau 
\sum_{s = \pm 1 } \sum_{i=0}^{N_s-1} \frac{U}{2} 
\bigl[ \hat{n}_{i0} (\tau) -  \alpha_{0}(s) \bigr ] \times\nonumber \\
&\bigl[ \hat{n}_{i1} (\tau) -  \alpha_{1}(s) \bigr ],
\end{align}
where
\begin{align}
&\begin{cases}
\alpha_{0}(s) = 1/2 + s \delta,\\
\alpha_{1}(s) = 1/2 - s  \delta\label{eq:alpha}
\end{cases}
\end{align}
where $\delta = 1/2 + 0^+$
 and $\hat{n}_{i\sigma} = \hat{c}_{i\sigma}^{\dagger} \hat{c}_{i\sigma}$.
The onsite Coulomb repulsion $U$ is assumed to be site-independent.
The parameters $\alpha_{\sigma}(s)$ are introduced \cite{Assaad:2007be,Rubtsov:2005iwa} to avoid a trivial sign problem.

On the other hand, in a single-orbital multi-site model with spin-orbit coupling, the presence of hopping terms that mix different spin flavors implies that $i$ and $j$ enumerate spin-sites ($N_s=2 N$) but $N_\sigma=1$.
In such cases, $i$ and $j$ enumerate spin-sites (the spin index runs first).
Accordingly, the interaction part is given by 
\begin{align}
	S_{\rm{int}} &= \int^{\beta}_{0}  d \tau 
	\sum_{s = \pm 1 } \sum_{i=0}^{N-1} \frac{U}{2} 
	\bigl[ \hat{n}_{2i,0} (\tau) -  \alpha_{0}(s) \bigr ] \times\nonumber \\
	&\bigl[ \hat{n}_{2i+1,0} (\tau) -  \alpha_{1}(s) \bigr ],
\end{align}
where $\alpha_{0}(s)$ and $\alpha_{1}(s)$ are the same 
as those defined in Eq.~(\ref{eq:alpha}).

The ALPS/CT-INT solver implements the continuous-time interaction-expansion QMC method~\cite{Rubtsov:2005iwa}.
A series expansion of the partition function is sampled in terms of $U$
using an efficient sampling method, the so-called submatrix update~\cite{Nukala:2009gh,Gull:2011hh,Nomura:2014by}.
For details of the submatrix updates in CT-INT refer to Ref.~\cite{Nomura:2014by}.

%

\section{Usage}\label{sec:usage}
\subsection{Requirements and installation}
The CT-INT code is built on an updated version of the core libraries of ALPS (Applications and Libraries for Physics Simulations libraries) [ALPSCore libraries]~\cite{Gaenko:2016ic}, the Boost libraries, and Eigen3.
Eigen3 is a \verb*#C++# template header-file-only library for linear algebra.
They must be pre-installed.
One needs a MPI \verb*#C++# compiler with support for \verb*#C++11# language features and CMake to build the solver.
It will install two executables ``ctint\_real" and ``ctint\_complex".
Only the difference between these two is that ctint\_real assumes the Weiss function to be real and runs faster in such cases.
The formats of input and output files are the same.


The latest version of the code is available from a public Git repository at \url{https://github.com/ALPSCore/CT-INT}.
One can also find a more detailed description of usage in Wiki documentation pages at
\url{https://github.com/ALPSCore/CT-INT/wiki}.

\subsection{Input data}
The essential input data of the solver are 
\begin{itemize}
	\item The complex-valued Weiss function defined on a grid in the interval $[0,\beta]$, 
	\item The onsite Coulomb interaction $U$.
\end{itemize}
We can also specify the number of thermalization steps, measurement steps, the interval of measurement of the Green's function.
All input parameters except for the Weiss function are read from a single input file.
The Weiss function must be stored in a separated text file in a given format.
The format of input files are described in the Wiki documentation pages.

\subsection{Execution}
Once you prepare two input files for runtime parameters and the Weiss function,
you can run the solver as follows.
\begin{lstlisting}
$ mpirun -np 120 ctint_real params.ini  
\end{lstlisting}
In this example, we specify the values of runtime parameters in params.ini.
The simulation results are written into a HDF5 file named params.out.h5.
Some examples are given in the following section.

\subsection{Output data}
The Green's function is defined as
\begin{align}
  G_{\sigma, ij} (\tau) \equiv - \braket{T_\tau\hat{c}_{i\sigma}(\tau) \hat{c}_{j\sigma}^\dagger(0)},
\end{align}
where $T_\tau$ is a time-ordering operator.
The Green's function is measured as
\begin{align}
  S_{\sigma, ij}(i\omega_n) &\equiv -\sum_k \Sigma_{\sigma, ik}(i\omega_n) G_{\sigma, kj}(i\omega_n).
\end{align}
In practice, the expansion coefficients in the Legendre representation  $(S_{\sigma, ij})_l$ are measured \cite{Boehnke:2011dd}.
The Matsubara-frequency data $S(i\omega_n)$ are reconstructed as
\begin{align}
  S_{\sigma, ij}(i\omega_n) &= \sum_{l=0} T_{nl} (S_{\sigma, ij})_l,
\end{align}
where $l$ is the index of Legendre polynomials.
The matrix elements $T_{nl}$ are introduced in Ref.~\cite{Boehnke:2011dd}.
Then, the Green's function can be reconstructed via the Dyson equation.
Equal-time quantities such as $\braket{n_{i\sigma} n_{j\sigma^\prime}}$ and $\braket{n_{i\sigma}}$ are also measured.

The results of the measurement are stored in a HDF5 \cite{hdf5} file.
The format of the output file is described in detail in Wiki documentation pages.

\section{Examples}\label{sec:examples}
\subsection{Three-site impurity model}
We consider a three-site model with onsite Coulomb repulsion.
The local Hamiltonian is given by
\begin{align}
\mathcal{H}&= - \sum_{i\neq j}^3 \sum_{\sigma} c^\dagger_{i\sigma} c_{j\sigma}
 - \mu \sum_{i=1}^3 n_{i\sigma} + U \sum_{i=1}^3 n_{i\uparrow} n_{i\downarrow},
\end{align}
where $c^\dagger_{i\sigma}$ and $c_{i\sigma}$ are 
creation/annihilation operators of an electron at site $i$ with spin $\sigma$.
The electron density operator is defined as $n_{i\sigma} \equiv c^\dagger_{i\sigma} c_{i\sigma}$.
For the bath, we use a semielliptical density of states with bandwidth equal to 4$t$, and set $t = 1$. 
This model is the same as that used for investigating a sign problem in a previous study~\cite{Shinaoka:2015wq}.

We solve the model for $U=4$ and $\beta=10$ at $\mu = U/2$.
The input file looks like this:
\begin{lstlisting}
total_steps  =  15000000
thermalization_steps  =  15000
measurement_period  =  10
model.beta  =  10.0
model.spins  =  2
model.U  =  4.0
model.sites  =  3
model.G0_tau_file  =  G0_TAU.txt
G1.n_matsubara  =  1000
G1.n_legendre  =  50
\end{lstlisting}
The data were obtain by running the solver with 120 MPI processes for 45 minutes.
The Green's function was measured every 10 Monte Carlo steps.
The program writes the following messages to the standard output at the end of the simulation:
\begin{lstlisting}
average matrix size was:
42.0968 42.0968
average sign was: 0.999426

#### Timing analysis ####
For measurement_period = 10 steps, each part took
Monte Carlo update: 194.746 ms
Recompute inverse matrix: 0.483439 ms
Global update: 0.000690577 ms
Measurement: 1.73871 ms
\end{lstlisting}

The average matrix sizes are around 42 for each spin.
The computational time spent for the measurement is negligible compared to that for the Monte Carlo updates.
This indicates that you could reduce the value of \textit{measurement\_period} to measure the Green's function more often.
The average sign is close to 1.

\begin{figure}
	\centering
	\begin{flushleft}\hspace{1em}(a)\end{flushleft}\vspace{-1em}
	\includegraphics[width=.3\textwidth,clip]{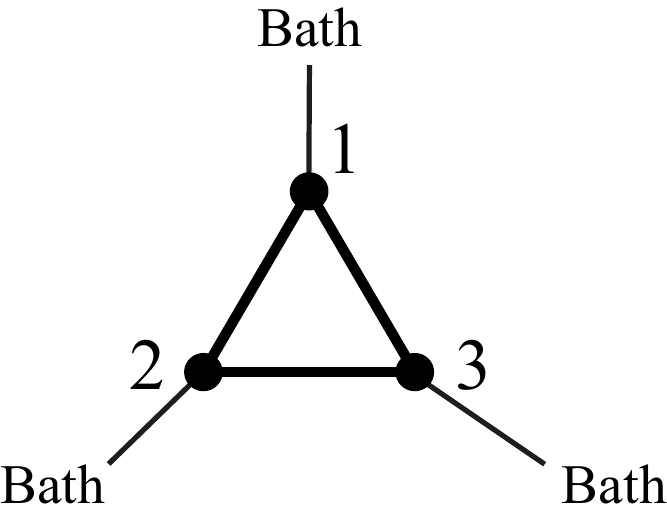}
	\begin{flushleft}\hspace{1em}(b)\end{flushleft}\vspace{-1em}
	\includegraphics[width=.45\textwidth,clip]{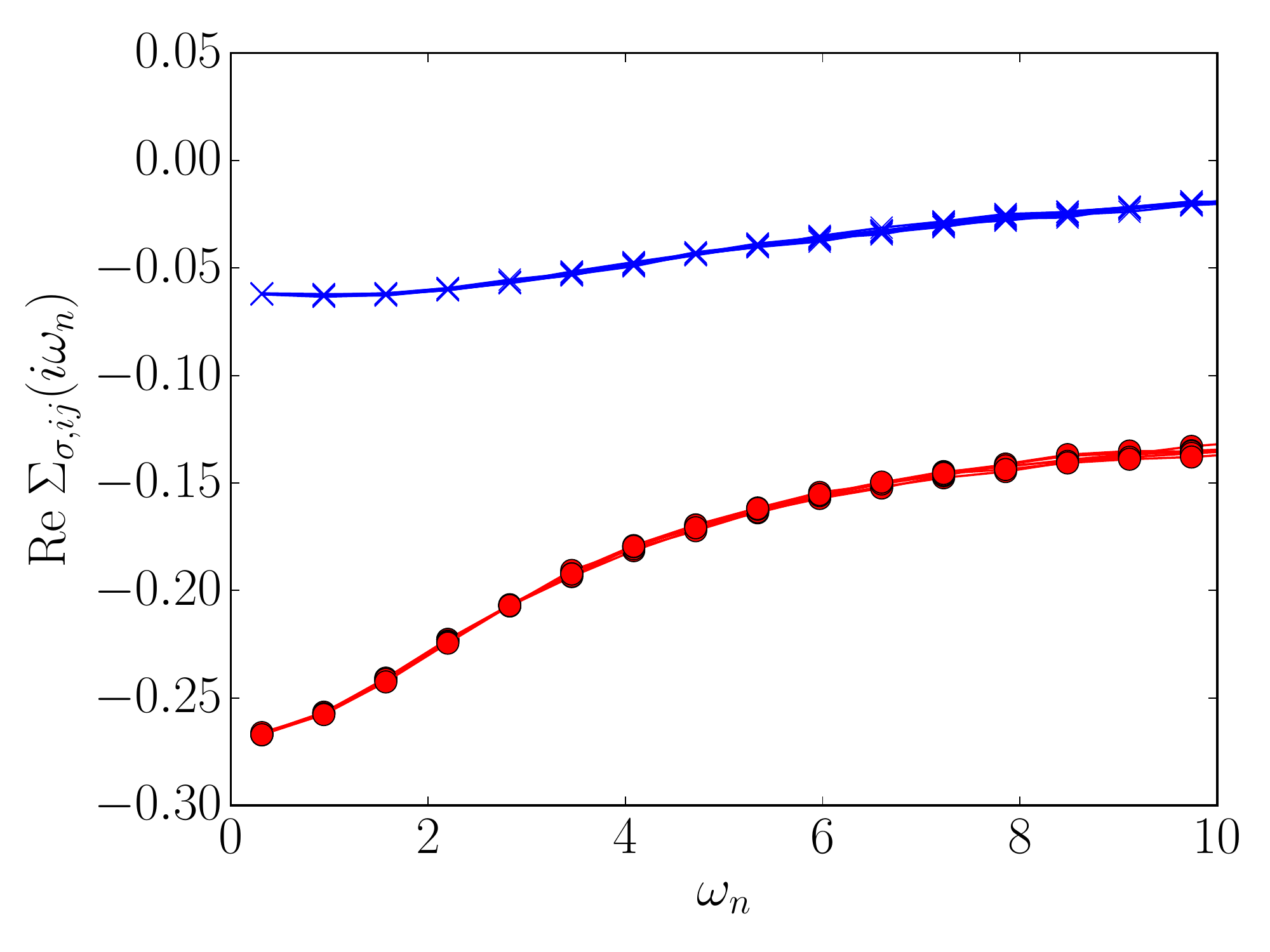}
	\begin{flushleft}\hspace{1em}(c)\end{flushleft}\vspace{-1em}
	\includegraphics[width=.45\textwidth,clip]{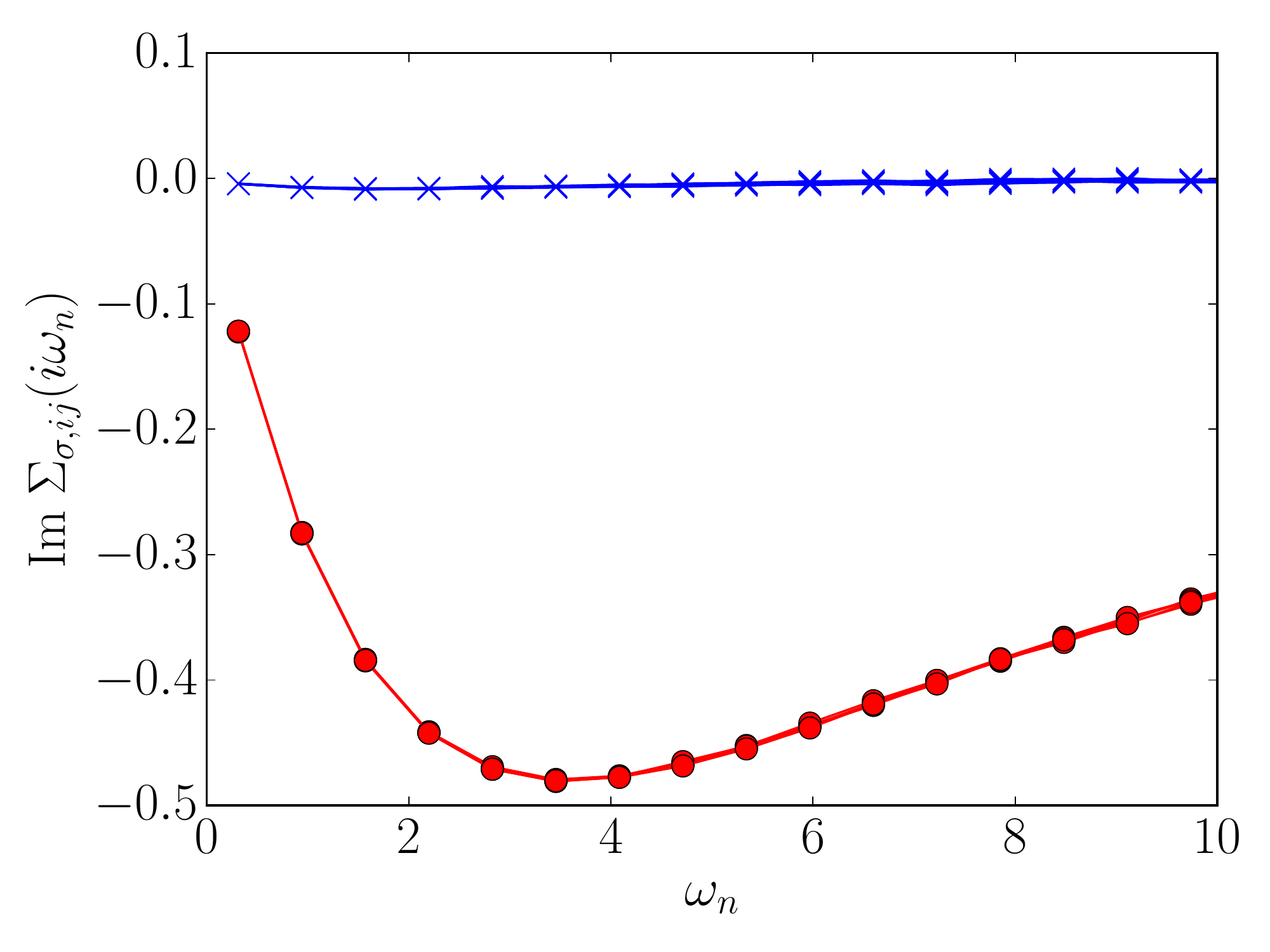}
	\caption{
		(Color online) 
		(a) Schematic of the three-site impurity model.
		Onsite [(b)] and offsite [(c)] components of $\Sigma_{\sigma,ij}(i\omega_n)$ computed for the three-site impurity model.
		The data for the onsite ($i=j$) and offsite ($i\neq j$) components are denoted by circles and crosses, respectively.
		We plot results for both of up and down spin components.
	}
	\label{fig:three-site}
\end{figure}

The computed results of $\Sigma_{\sigma,ij}(i\omega_n)$ are shown in Fig.~\ref{fig:three-site}.
The self-energy was obtained by solving the Dyson equation.
The data for up and down spins agree within error bars.

\subsection{Self-consistent calculations of the 2D Hubbard model within the dynamical cluster approximation}
The ALPS/CT-INT can be used together with the DCA for solving the 2D Hubbard model on the square lattice.
The Hubbard model on the 2D square lattice is defined as 
\begin{align}
\mathcal{H}= & -t  \sum_{ \langle i,j \rangle}  \sum_{\sigma} c^\dagger_{i\sigma} c_{j\sigma}
 -t'  \sum_{ \langle  \langle i,j \rangle \rangle}  \sum_{\sigma} c^\dagger_{i\sigma} c_{j\sigma}  \nonumber \\ 
 & + U \sum_{i} n_{i\uparrow} n_{i\downarrow}, 
\end{align}
where $\langle i,j \rangle$ and $\langle\langle i,j \rangle\rangle $ indicate pairs of nearest neighbor and next nearest neighbor sites, respectively.
We take nearest neighbor hopping amplitude $t$ as energy unit, i.e., $t=1$. 
$t'$ and $U$ are the next nearest neighbor hopping and Hubbard interaction, respectively. 

\begin{figure}
  \centering\includegraphics[width=.3\textwidth,clip]{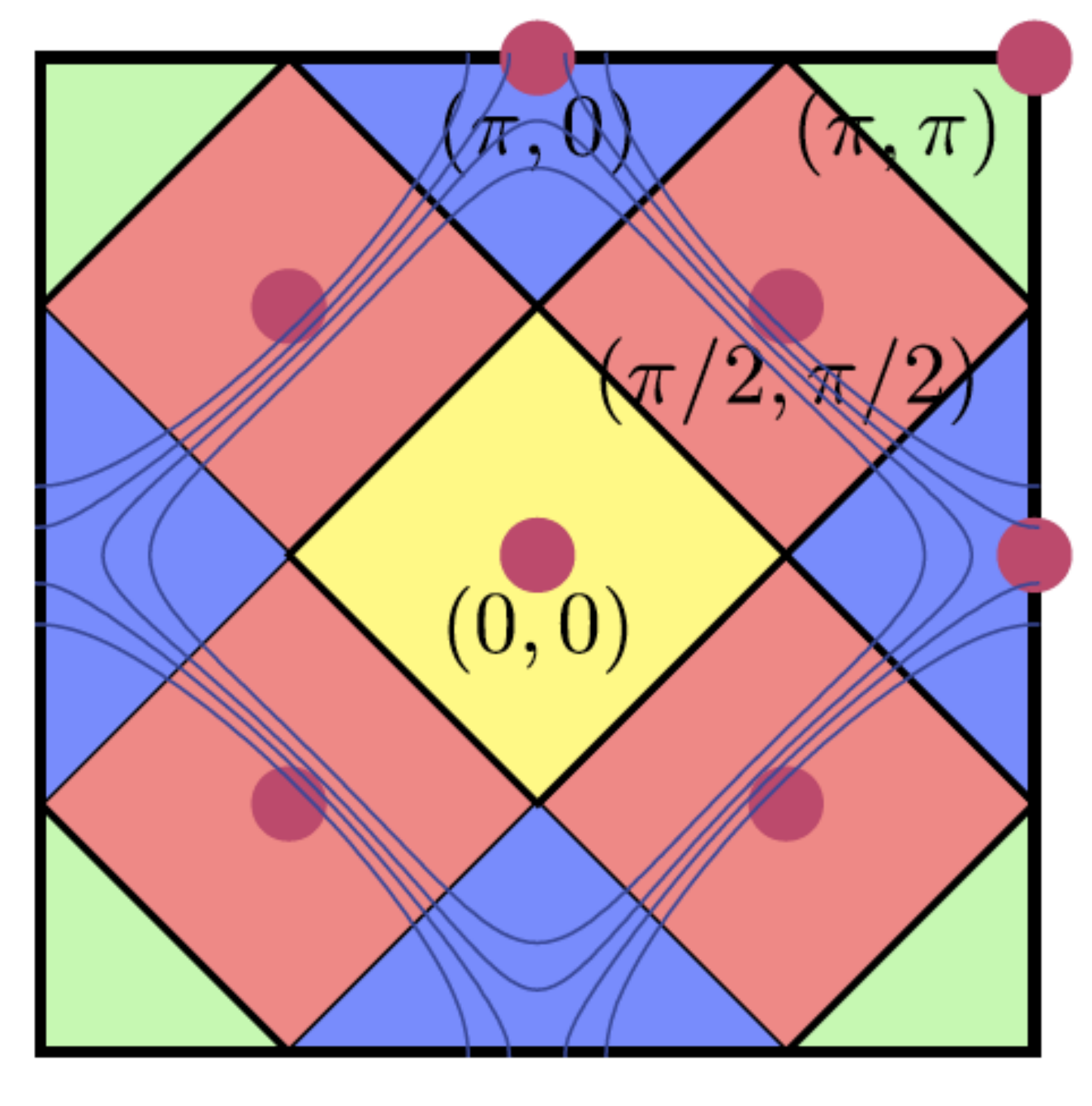}
	\caption{
		(Color online) 
		Momentum patches used in the 8-site DCA \cite{Maier:2005et} calculation. 
		Thin lines indicate the fermi surfaces of noninteracting system ($t'=-0.15$) at half-filling and 10, 20, 30 \% hole dopings. 
		Dots represent central momenta to specify patches. 
		Adapted from Ref.~\cite{Gull:2010by}.
	}
	\label{fig:dca_geometry}
\end{figure}

Here we employ 8 site DCA, in which the Brillouin zone is partitioned into 8 patches (Fig.~\ref{fig:dca_geometry}). 
Each patch is labeled by the central momentum ${\mathbf K}$. 
Due to symmetry, the number of inequivalent patches is reduced to 4: 
The inequivalent patches are $(0,0)$, $(\pi/2,\pi,2)$, $(\pi,0)$, and $(\pi,\pi)$. 
In each patch, the self-energy is approximated by that of central momentum i.e.,  $\Sigma({\mathbf k},i\omega_n) = \Sigma_{{ \mathbf K}} (i\omega_n)$.
We also assume a nonmagnetic solution,
dropping the spin index hereafter.
Within this approximation, the Green's function for each patch is given by 
\begin{align}
 G_{{ \mathbf K}} (i\omega_n)  = \int^{{ \mathbf K}}  \! \! d{\mathbf k} \  \frac{1}{i\omega_n + \mu  - \epsilon_{\mathbf k }  -  \Sigma_{{ \mathbf K}} (i\omega_n) }
\end{align}
where the integral is done at each momentum patch. 
$\epsilon_{\mathbf k }$ is the energy dispersion of noninteracting system given by $\epsilon_{\mathbf k } = -2t (\cos(k_x) + \cos(k_y)) - 4t' \cos(k_x)\cos(k_y)$. 
Further details of the self-consistency can be found in Ref. \cite{Maier:2005et}.

\begin{figure}
	\centering
	\includegraphics[width=.5\textwidth,clip]{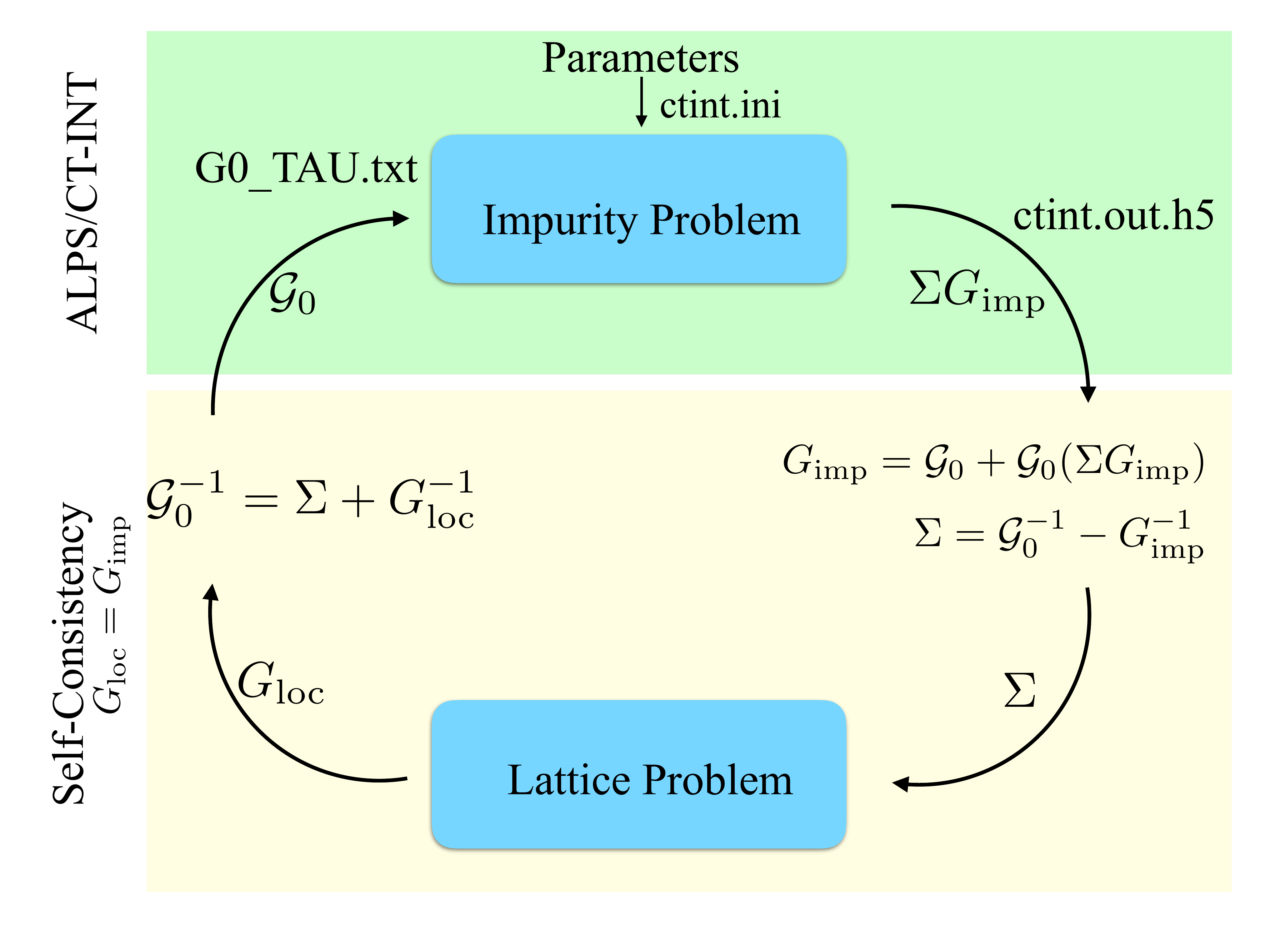}
	\caption{
		(Color online) Illustration of the self-consistent cycle in DCA.
		The names of the input and output files of ALPS/CT-INT can be specified at runtime.
	}
	\label{fig:DCA_cycle}
\end{figure}

\begin{figure}
    \begin{flushleft}\hspace{1em}(a)\end{flushleft}\vspace{-1em}
    \centering\includegraphics[width=.45\textwidth,clip]{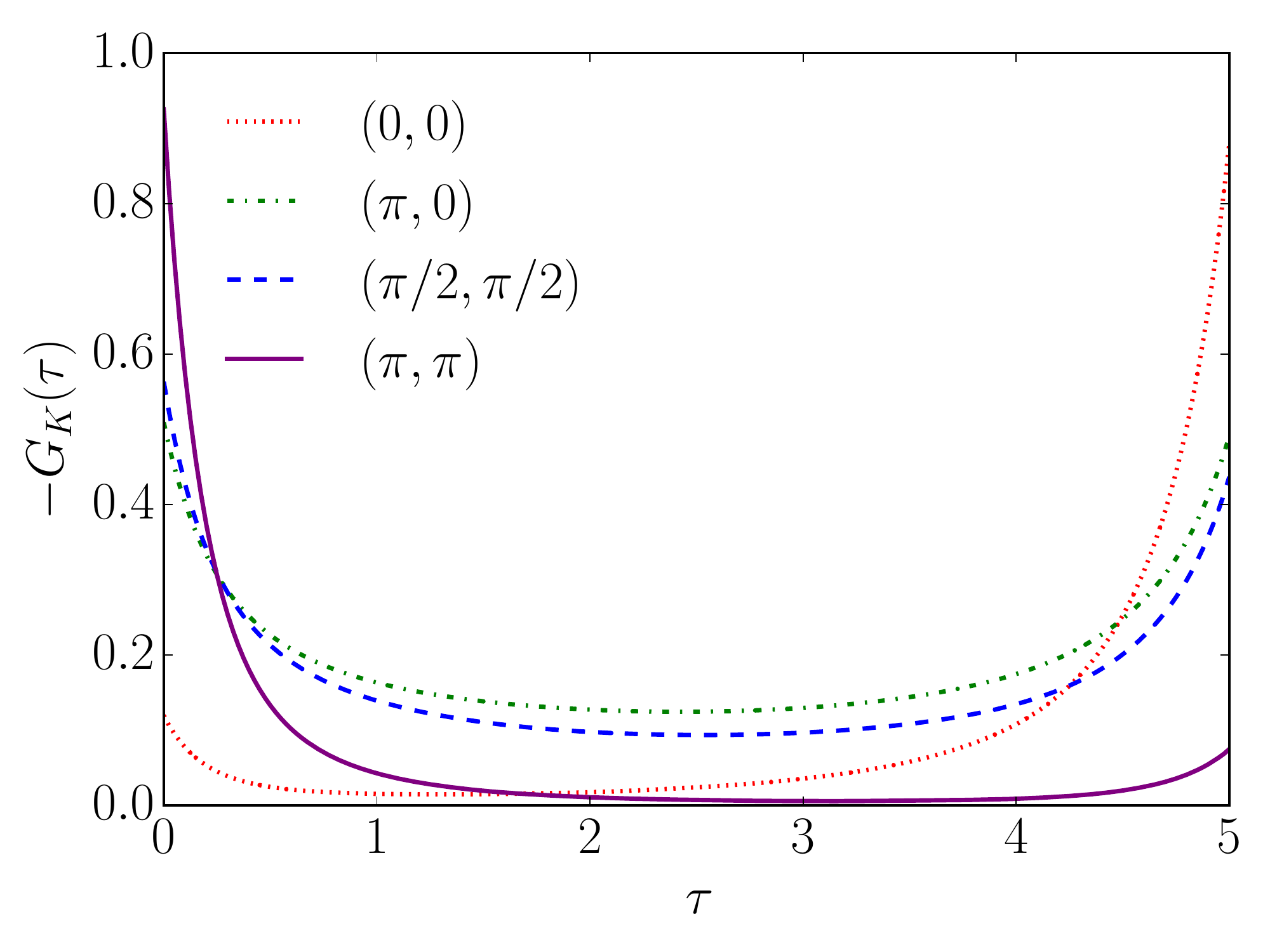}
    \begin{flushleft}\hspace{1em}(b)\end{flushleft}\vspace{-1em}
    \centering\includegraphics[width=.45\textwidth,clip]{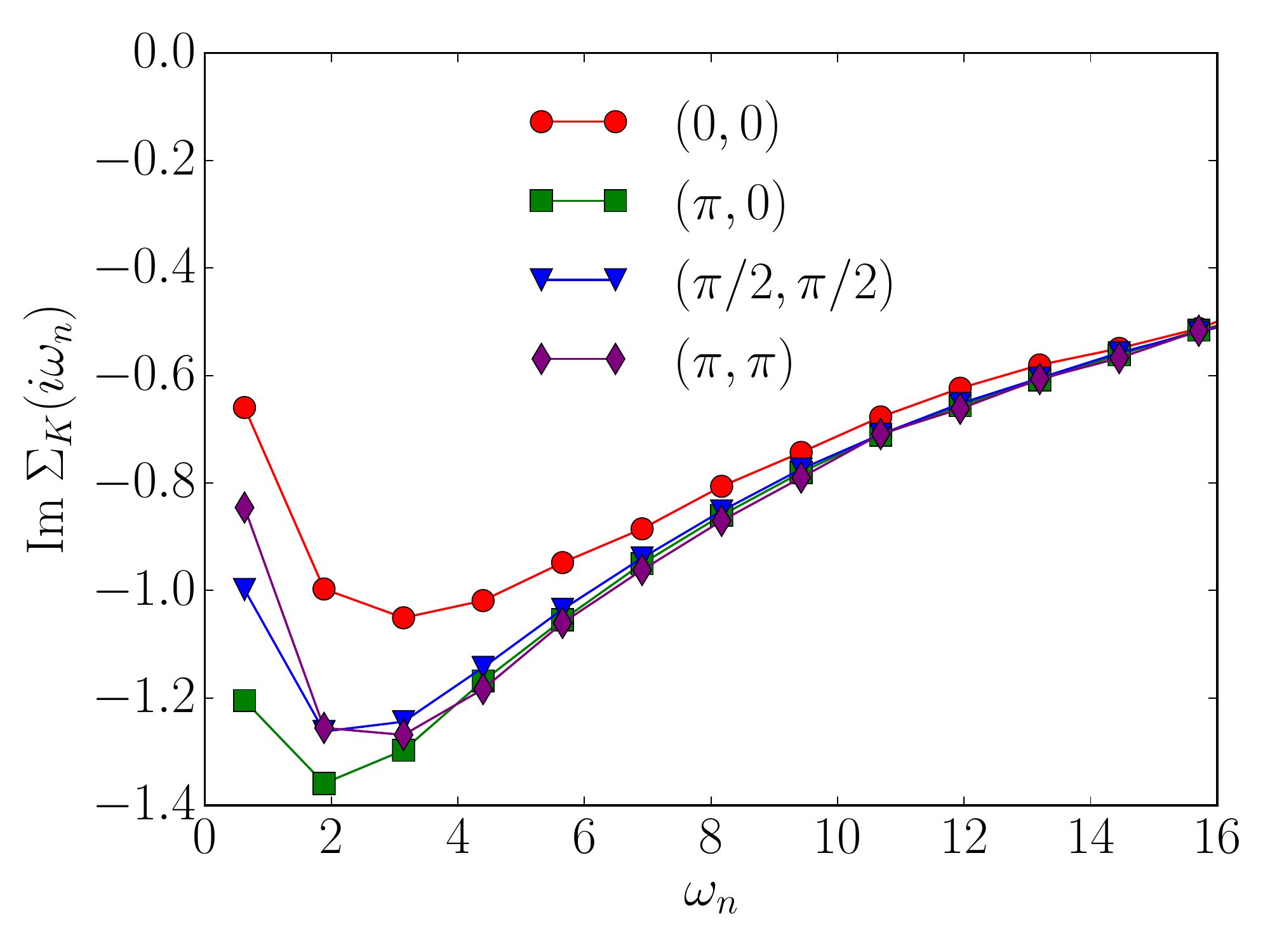}
	\caption{
		(Color online) 
		8 site DCA results of (a) Green's function and (b) the imaginary part of self-energy for $t^\prime=-0.15$, $U=6$, $\mu = U/2 - 1.0$, and $\beta=5$. 
	}
	\label{fig:dca_result}
\end{figure}

We perform self-consistent calculations combining the ALPS/CT-INT solver and an external program for solving self-consistent equations.
Figure~\ref{fig:DCA_cycle} illustrates the self-consistent cycle of DCA. 
An input file for ALPS/CT-INT, ``ctint.ini" looks like this:
\begin{lstlisting}
model.beta  =  5.0
model.spins  =  2
G1.n_matsubara  =  1000
total_steps  =  600000
G1.n_legendre  =  60
model.U  =  6.0
measurement_period  =  10
thermalization_steps  =  500
model.sites  =  8
model.G0_tau_file  =  G0_TAU.txt
\end{lstlisting}

Figure~\ref{fig:dca_result} shows the 8 site DCA results of (a) Green's function and (b) the imaginary part of self-energy for $t'=-0.15$, $U=6$, $\mu = U/2 - 1.0$, and $\beta=5$.

\section{Summary}\label{sec:summary}
We have presented an open-source \verb*#C++# implementation of the continuous-time interaction expansion Monte Carlo method for impurity models with certain types of density-density Coulomb interactions and general hybridization functions.
More general forms of interaction will be supported in a future version.
We have discussed the technical details of the implementation.
We presented some examples of Monte Carlo simulation results for a three-site model as well as results of 8-site DCA calculations for the two-dimensional Hubbard model.
They can serve as a benchmark or reference. 

\section*{Acknowledgments}
We gratefully acknowledge support by the wider ALPS community~\cite{Bauer:2011tz,Albuquerque:2007ja}.
HS was supported by JSPS KAKENHI Grant No. 16H01064 (J-Physics), 18H04301 (J-Physics), 16K17735.
YN was supported by Grant-in-Aids for Scientific Research (JSPS KAKENHI) Grant No. 17K14336 and 16H06345.
HS and YS were supported by JSPS KAKENHI Grant No. 18H01158.
EG was supported by NSF DMR 1606348.
Part of the calculations were performed on the ISSP supercomputing system.
 
\section*{References}
\bibliographystyle{elsarticle-num}

\appendix
\end{document}